# Quantum Tunneling Hygrometer with Temperature Stabilized Nanometer Gap

A. Banerjee (A.B), R. Likhite (R.L), H.Kim (H.K) and C.H. Mastrangelo (C.M)*

University of Utah, 201 Presidents Cir, Salt Lake City, UT 84112

**Introduction**

We present the design, fabrication and response of a humidity sensor based on quantum tunneling of electrons through temperature-stabilized nanometer gaps. The sensor consists of two stacked metal electrodes separated by ~2.5 nm of vertical air gap. Upper and lower electrodes rest on separate 1.5 μm thick polyimide patches with nearly identical thermal expansion but different gas absorption characteristics. When exposed to a humidity change, the patch under the bottom electrode swells but the patch under the top electrode does not, as it is covered with a water-vapor diffusion barrier ~8 nm of $Al_2O_3$. The air gap thus decreases leading to increase in the tunneling current across the junction. The gap however is independent of temperature fluctuations as both patches expand or contract by near equal amounts. Humidity sensor action demonstrates an unassisted reversible resistance reduction $R_{max}/R_{min}$ ~$10^5$ when the device is exposed to 20 – 90 RH% at a standby DC power consumption of ~0.4 pW. The observed resistance change when subject to a temperature sweep of 25 – 60° C @24% RH was ~0.0025% of the full device output range.

**Introduction**

Humidity sensors can be realized using various technologies [1]–[5]. Kharaz and Jones were one of the first to report optical methods of humidity measurement [6]. First introduced by Wohltjen in 1984, SAW sensors were widely used as hygrometers, where shift in resonant frequencies was a direct indication of mass of adsorbed water vapor [7]. The two most common type of hygrometer devices are resistive and capacitive type.

Capacitive humidity sensors are the most common realizations of microfabricated hygrometers [8]. These devices are used in approximately 75% of applications [9]. In these devices, the dielectric properties of a hygroscopic layer (usually a polymer such as polyimide) change when exposed to humidity. This change may also be accompanied by swelling of the hygroscopic layer. Micro-cantilevers have also been used as capacitive type humidity sensors. In these devices, a thin polyimide film is patterned on the cantilever and when exposed to humidity, the polymer film tends to expand thereby exerting mechanical stress on the cantilever causing it to bend [9]. Capacitive humidity sensors display linear operation and can accurately measure from 0% -100% relative humidity with an output range typically ~30% of the default capacitance. These devices consume very little electrical power and are relatively cheap to fabricate [3].

Resistive humidity sensors are also extensively used. In these devices the resistance of the sensing material changes when exposed to humidity. Ceramics such as $Al_2O_3$ [10], conductive polymers such as PVA with graphitized carbon [11] and polyelectrolytes [12] have been used as functional materials for resistive humidity sensors. In contrast to capacitive devices, resistive humidity sensors are highly nonlinear and capable of providing several orders of magnitude change in resistance. An advantage of resistive RH sensors is their interchangeability, typically within ±2% RH. Operating temperatures of resistive sensors ranges from –40°C to 100°C; however these devices have significant temperature dependencies. For example, a typical temperature coefficient for a resistive RH sensor is -1% RH / °C.

In this work we present a new type of microfabricated humidity sensor that is able to provide large output range and a low temperature dependence. The device utilizes the expansion of a polymer that swells when exposed to humidity as in the capacitive device, but it produces a resistive output that measures the polymer expansion through tunneling current across a humidity dependent, thermally stabilized nanogap. The tunneling current changes many orders of magnitude providing similar output as the resistive type device with a low temperature dependence.

In the first part of the article, we focus on the fabrication and electrical characterization of the device. The second part of the paper will describe the device response to change in ambient RH% and temperature response of the device. The third part of the article deals with relevant analysis of the current conduction across the nanogap and a mathematical model describing current flow as well as an equivalent electrical circuit. This section also includes a discussion of the sorption kinetics of the device and analysis of the absorption-desorption phenomenon

**Basic Operating Principle of Resistive Nanogap Hygrometer**

Nanogap electrodes (electrodes separated by a gap of a few nanometers), have been previously used for highly sensitive bio-sensor applications [13][14][15][16][17][18][19]. Similar to other devices built by Mastrangelo et al [20][21][22][22][23][24][23], these can be used as low-power and highly sensitive tunneling hygrometers for IoT applications. In this article, we show the working of a new type of tunneling hygrometer. The device consists of a pair of upper and lower electrodes, separated by an air-gap of ~2.5 nm as shown in Fig. 1(a) below illustrating the basic sense mechanism.

The upper electrode rests at a fixed height. The bottom electrode rests on top of a hygroscopic polymer, polyimide that swells when humidity is absorbed; thus reducing the gap between the two electrodes. The swelling of the polyimide with humidity is linear corresponding to its humidity coefficient of expansion or CHE, which is 60-75 ppm/% RH [25]. Fig. 1(b) shows the electron band diagram across the nanogap. If the nanogap is very small electrons can tunnel from one electrode to the other establishing a conductive path. The magnitude of the tunneling current is exponentially dependent on the gap [26] hence the device resistance goes as

$$R(RH\%) = R_o \cdot e^{(a*\Delta(RH\%))} \sim A(T) \cdot e^{(-B \cdot (RH\%))} \qquad (1)$$

Where *A(T)* and *B* are fitting parameters. In the simple configuration shown in Fig. 1(a), the gap *Δ* is also affected by the thermal expansion of the polyimide, which makes the coefficient *A* and the resistance of the device strongly dependent of the ambient temperature.

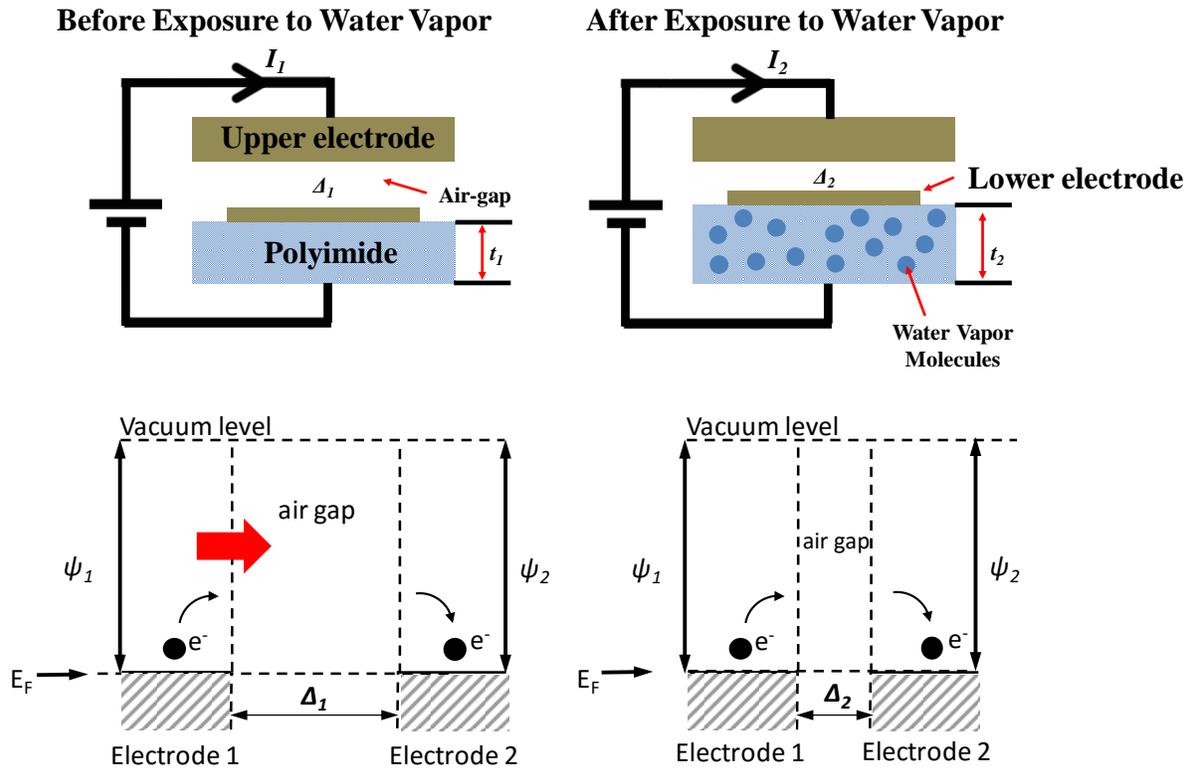

**Figure 1**. Working principle of the tunneling humidity sensor. Exposure to water vapor molecules reduces the gap and reduces tunneling distance.

To eliminate this strong temperature dependence we utilize the differential device arrangement as shown in the schematic of Fig. 2 below where the two electrodes rest on polyimide patches of equal thickness, hence in the absence of humidity the nanogap distance is constant and independent of temperature fluctuations. However only one of these patches can absorb humidity; thus producing a humidity induced nanogap change. The differential device assembly consists of an upper Al electrode and a lower Cr electrode separated by an air-gap of ~2.5 nm, standing on separate patches of 1.5 μm thick polyimide. As shown in Fig. 2, the polyimide patch under the upper electrode is covered with ~8 nm of ALD deposited $Al_2O_3$ diffusion barrier whereas that under the lower electrode is exposed to ambient atmosphere.

We thus achieve nanogap temperature stabilization by using cancelation of a common mode thermal expansion of both patches and humidity signal extraction by differential response to humidity between the two patches. Since both the top and bottom electrodes are standing on near identical polyimide patches, any increase in ambient temperature would lead to both the patches expanding almost equally. This ensures that in the event of temperature fluctuation, the nanogap distance between the top and bottom electrode will remain unchanged and the electrical response will be negligible in comparison to sensor response.

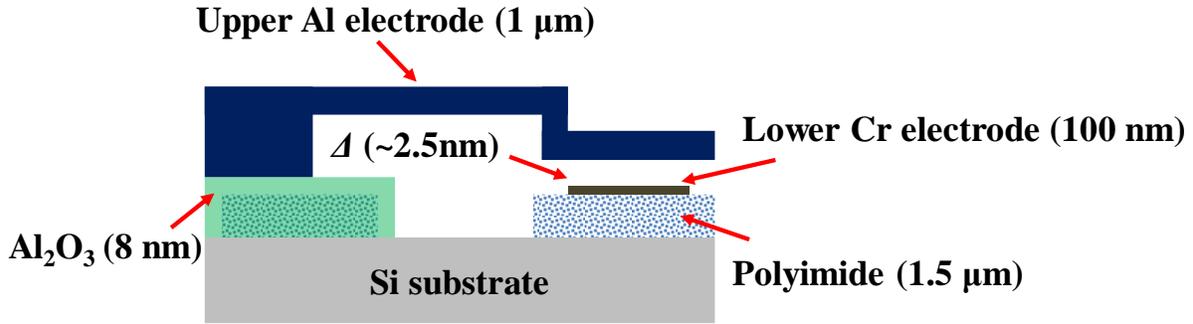

**Figure 2**. Schematic of the fabricated device.

**Results and Discussion**

*Fabrication and Electrical Characterization*

The simplified fabrication process is shown in Figure 3. A more detailed description of the fabrication process is given in the "methods" section. Figure 4a shows high resolution SEM images of the fabricated sensor and Figure 4 b shows *I-V* characteristics of the device after fabrication of the device. The junction *I-V* characteristics are typical of very low current on account of very low electron tunneling across an air-gap of ~2.5 nm. The average DC resistance of the junction is measured to be ~271 GΩ. This shows an extremely low leakage current under low biasing conditions.

*Sensor Action and tunneling current at various relative humidity levels*

After the device is fabricated, the upper and lower electrodes are separated by an air-gap of ~ 2.5 nm. Therefore current flow across the nanogap junction involves conducting electrons tunneling through 2.5 nm of air-gap. However, when the device is exposed to an increase in humidity, only the un-protected patch of polyimide which is beneath the lower electrode absorbs water-vapor molecules and swells. This is because the $Al_2O_3$ acts as a diffusion barrier and prevents the polyimide patch under the upper electrode from absorbing water molecules [27]. This differential swelling of the polyimide patches results in the inter-electrode distance to reduce below 2.5 nm. Therefore, after absorption of water vapor molecules, the tunneling distance for the conduction electrons reduces. Since tunneling current exhibits exponential dependence on tunneling distance, increase in humidity levels lead to an exponential increase in tunneling current. In other words, junction resistance decreases exponentially when exposed to increased humidity levels.

Figure 5a shows the I-V curves of the device when exposed to an increasing RH% from ~20 – 90 RH%. As evident from the *I-V* plots, the increasing RH% leads to an increase in magnitude of current flow for the same voltage bias. Figure 5b is a plot of normalized average resistance vs RH%. These plots are a clear indication of exponential reduction of junction resistance with increasing RH%. Figure 6 a-c shows five cycles of repetitive exposure of device to an increase in ambient RH% from 20-90 for three separate devices, and subsequent removal of excess humidity from the testing chamber. As evident from the plot, the device exhibits unassisted and perfect recovery of junction characteristics after removal of water vapor from the testing chamber. The plots also suggest that there is a difference in time taken to reach the maximum resistance drop for different devices. The reason is simply because of the fact that the time taken for the chamber to reach the same humidity level varied between successive testing cycles and during multiple device testing. Our fabricated device followed the commercially available BME280 reference sensor perfectly. Any delay in sensor response was also exhibited by the reference chip as well. Figure 7 shows the comparative plots of successive cycles of moisture absorption and desorption by the device. As evident from the plot, the device doesn't suffer from moisture hysteresis or sensor saturation.

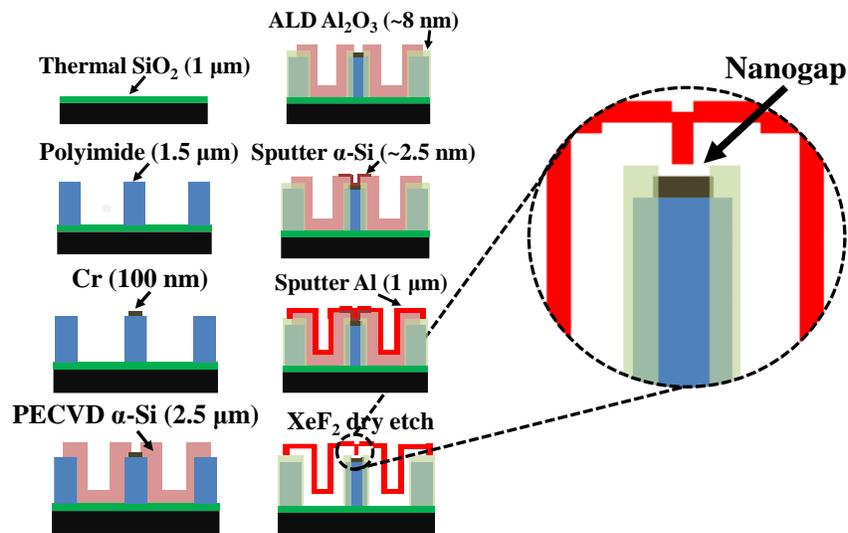

**Figure 3**. Simplified fabrication flow and zoomed in view of the nanogap junction.

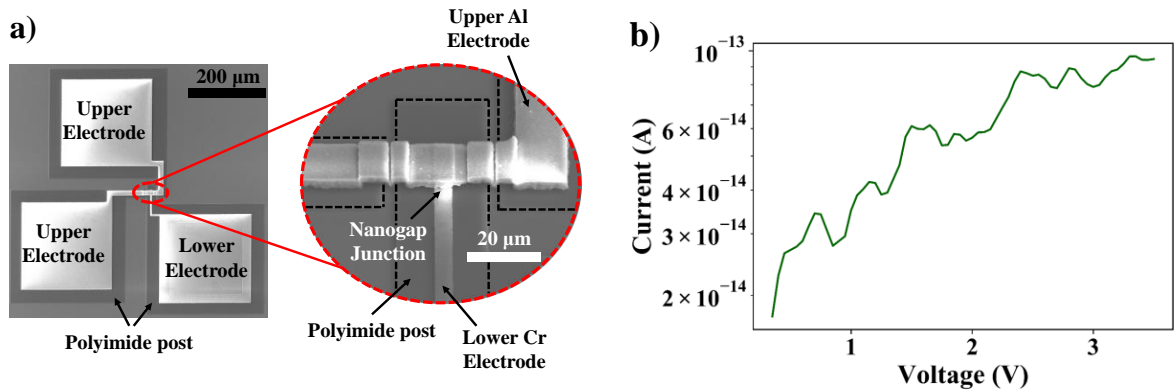

**Figure 4.** a) SEM images of fabricated device b) I-V characteristics across the tunneling junction after sacrificial etching of the α-Si.

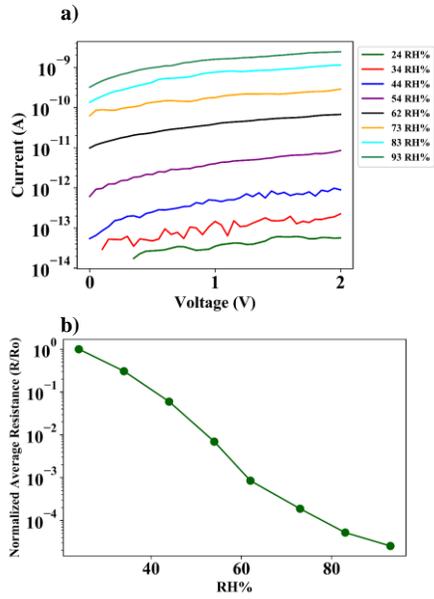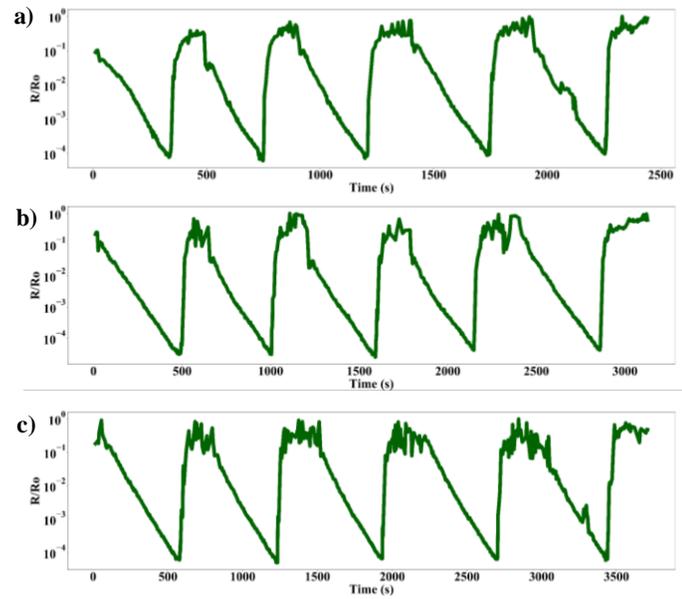

**Figure 5.** a) *I-V* characteristics of device b) Normalized average junction resistance at various humidity levels in the test chamber.

**Figure 6.** Repetitive cycles of water vapor exposure and removal from testing chamber for three different devices.

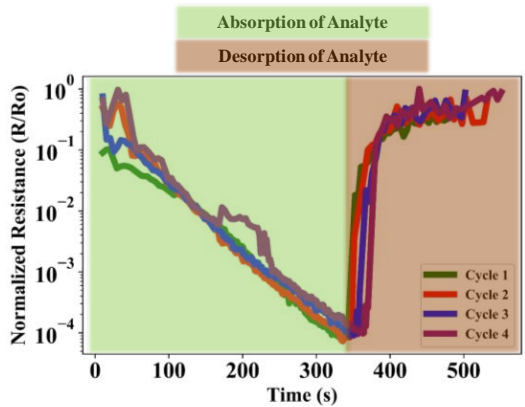

**Figure 7.** Comparison of sensor response for multiple cycles of moisture exposure and removal for the same device.

*Temperature response of the device and passive temperature compensation*

Researchers have always tried to neutralize the undesired temperature response of polymer based sensors [9]. Essentially, temperature compensation can be achieved using passive and active methods. In passive temperature compensation, passive electrical elements such as resistors are employed to compensate pre-determined fluctuations in sensor readout due to temperature changes. Active temperature compensation involves an active temperature feedback to the transducer's signal processing circuit in order to compensate for the temperature drift. Recently, Rinaldi [28] and Mastrangelo [22] used self-levelling beams in order to achieve temperature compensation in micromachined sensors. In our proposed tunneling hygrometer, we achieve passive temperature compensation by using a non-differential polymer expansion design when exposed to temperature fluctuations. As shown in Figure 8, when the device is exposed to water vapor molecules, the polymer platform which does not have a protective layer of $Al_2O_3$, absorbs the molecules and expands, but the protected polymer patch does not. Therefore, air-gap between the electrodes reduces and tunneling current flow between them increases. However, when the device is subjected to an increase in temperature, due to near identical thermal mass and negligible thickness of the $Al_2O_3$ layer in comparison with the platform, both polyimide patches expand equally. This ensures that the air-gap between the electrodes remains almost constant.

Figure 9 shows the temperature response of the device when subjected to a temperature change from 25-60 °C. As evident from the plot, the junction reduces <5 times when subjected to temperature changes. This is ~0.05% of the maximum resistance drop of the junction resistance when exposed to change in humidity. Therefore, it is clear that the device displays sufficient temperature compensation.

*Current conduction mechanism, mathematical model and equivalent electrical circuit*

The current conduction through the nanogap junction is considered to be electron tunneling. Essentially, current flow across the air-gap is possible only when conduction electrons tunnel between the metal electrodes. As explained above, when the device is exposed to an increasing RH%, the gap between the electrodes reduce. Tunneling current exponentially depends on the distance through which the electrons have to tunnel through. Therefore, exposure to rising levels of humidity leads to an exponential reduction of junction resistance. The air-gap between the upper and lower electrode can then be considered as a function of RH% and coefficient of hygroscopic expansion of polyimide. The current conduction can be modelled by using the generalized Simmons' expression for electron tunneling current density [29]. Figure 10a shows the equivalent electrical circuit describing our sensor action.

For intermediate voltages, current flowing through the nanogap junction can be written as:

$$I_T(V, C_g) = (G_{SO}) \cdot \left[ \frac{\Phi_a}{q} \cdot e^{\left(-2 \cdot ((d_s - \beta \cdot t_p \cdot C_g) \cdot \sqrt{\frac{2m}{\hbar^2} \cdot \Phi_a}\right)} + \left(\frac{\Phi_a}{q} + V\right) \cdot e^{\left(-2 \cdot (d_s - \beta \cdot t_p \cdot C_g) \cdot \sqrt{\frac{2m}{\hbar^2} \cdot (\Phi_a + qV)}\right)} \right] \quad (2)$$

Where, $I_T(V, C_g)$ = current across the air-gap as a function of applied voltage and relative humidity of the testing chamber, $V$ = Biasing voltage across device, $C_g$ = Relative humidity of the testing chamber, $\beta$ is a fitting parameter proportional to the coefficient of hygroscopic expansion for polyimide, $t_p$ = thickness of polyimide platform, $\Phi_a$ is the mean barrier potential of the air-gap, $d_s$ = original air-gap, $m$ = effective mass of tunneling electrons, $\hbar$ = reduced Planck's constant, $q$ = charge of electron and $G_{SO}$ is a conductance fitting factor proportional to the overlap area and having units of S.

Figure 10b shows the *I-V* plots of the device after being exposed to varying levels of relative humidity curve fitted with Equation 1. As evident from the plot, the experimental data is in good agreement with the mathematical model. Parameter extraction revealed the average value of $\beta$ to be ~2.78 ppm/RH% and $G_{SO}$ to vary from $5.2 \times 10^{-4}$ $\Omega^{-1}$ for the *I-V* curve corresponding to 24 RH% to ~ 10 $\Omega^{-1}$ for the *I-V* curve corresponding to 83 RH%. The root-mean-square-error of the curve-fitting plots shown in figure 10b was found to be 10%, 13%, 1.5% and 6% of the average experimental data of the I-V curves corresponding to junction electrical characteristics when the sensor was exposed to 24 RH%, 44 RH%, 62 RH% and 83 RH% respectively.

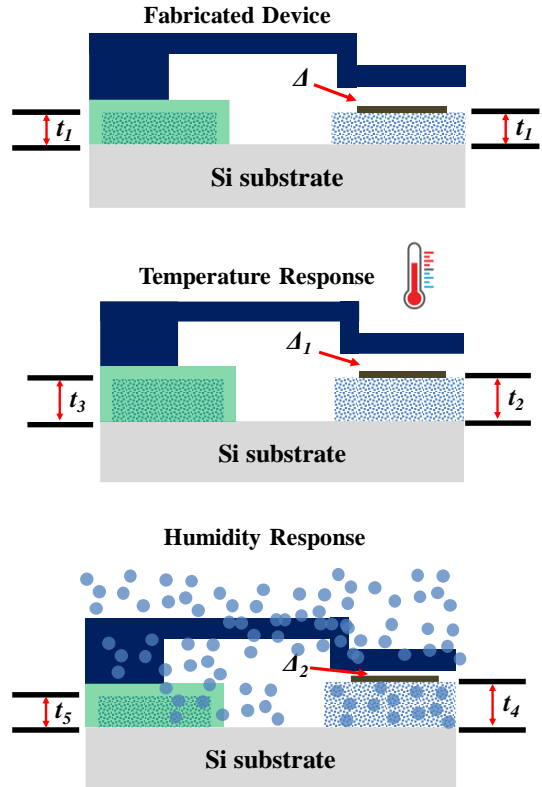

**Figure 8.** Schematic of non-differential swelling design to achieve passive temperature compensation.

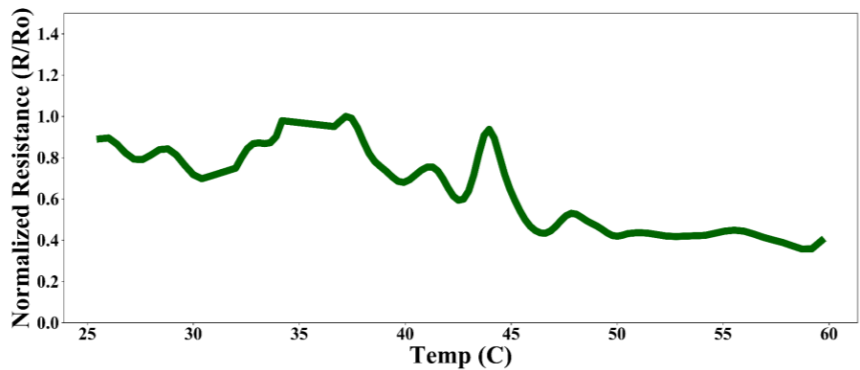

**Figure 9**. Temperature response of the device.

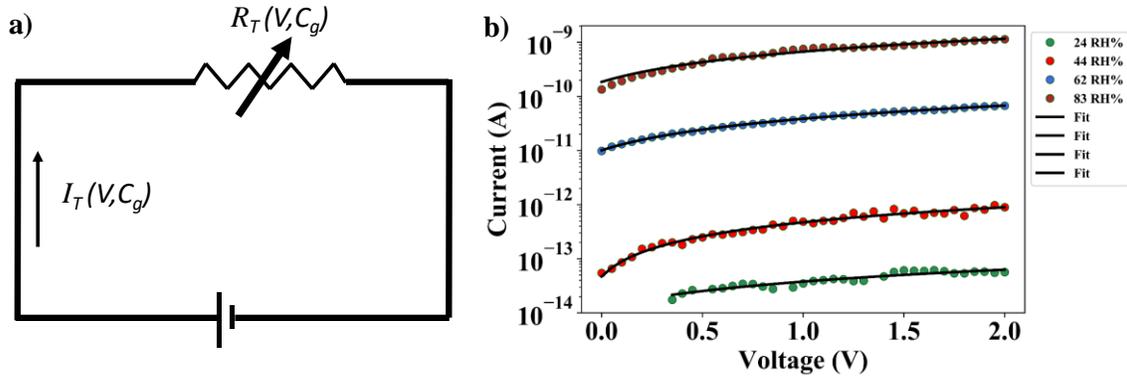

**Figure 10.** a) Equivalent electrical circuit of the sensor b) Curve-fitted plots of the I-V measurements of the device after exposure to different levels of moisture content in the test chamber.

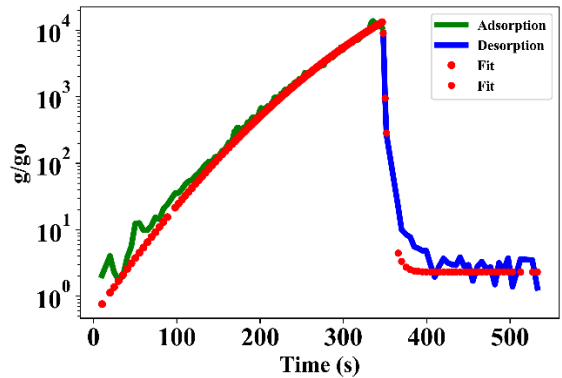

**Figure 11**. Normalized conductance response to one period of increasing exposure of humidity levels to nanogap device and subsequent removal, curve-fitted to Fick's law of diffusion and Polanyi-Wigner desorption model respectively.

*Absorption-Desorption dynamics*

Since sensor dynamics is heavily dependent on the diffusion of water vapor molecules into the polyimide patch causing expansion of the polymer, the dynamic response of the absorption cycle of the sensor can therefore by mathematically described using a modified version of Fick's second law [30]. Fick's law of diffusion has been extensively used to describe the diffusion process and to experimentally determine diffusion constant of various substances. The desorption process ideally consists of excess water molecules in the polyimide desorbing back into the atmosphere into water-vapor molecules. This typically is characteristic of a simple first-order desorption process where the absorbed gas-molecule on the surface of the solid simply desorbs back into its gaseous form. The kinetic process is mathematically described by the Polanyi-Wigner equation and is used as the theoretical basis for thermal desorption spectroscopy process [31]. Figure 11 shows the normalized conductance of the sensor as a function of time for one cycle of water-vapor absorption and desorption, where the absorption cycle is curve-fitted with Fick's law of diffusion and the desorption cycle is curve-fitted with the Polanyi-Wigner desorption model. As evident from the plot, the presented mathematical model accurately describes the sensor action.

*Methods*

Device Fabrication The fabrication process started by growing ~1 μm of thermal $SiO_2$ on a 4-inch Si wafer. Polyimide was then diluted by dissolving uncured HD4104 polyimide (purchased from HD Microsystems) in N-Methyl-2-pyrrolidone (NMP) solvent in the ratio of 1:0.5 (wt/wt). This mixture was then dissolved by using a stirring it with a magnetic stirrer at 300 rpm for 2 hours to ensure perfect mixing of the polyimide and the NMP solvent. This mixture was then spin-coated on the sample at 2000 rpm following the standard procedures for polyimide processing. For curing the polyimide, the sample was kept in a nitrogenized environment oven for three hours at a temperature of 300 ° C. This procedure resulted in polyimide thickness of ~1.5 μm. Following this, we sputtered 200 nm of Al on the sample at 50 W and used it as a hard-mask to pattern the underlying polyimide. $O_2$ plasma dry etching for 10 minutes at 100W was sufficient to remove the unwanted polyimide from the sample. Next, the remnant Al was stripped off by using commercially available aluminum etchant. Following this, we sputtered ~ 100 nm of Cr at 50 W on the sample and lithographically patterned it to define the lower electrodes. After this, we deposited about 2.5 μm of PECVD α-Si on the sample to form the sacrificial bridging supports for the upper electrode. This was followed by thermal ALD of ~8 nm $Al_2O_3$ and its lithographic patterning using BOE. Then, we sputtered ~2.5 nm of α-Si on the sample at 50 W to pattern the sacrificial spacer layer to define the thickness of the eventual nanogap between the electrodes. We next deposited ~1 μm of Al on the sample and patterned it to form the upper electrode. Finally, we sacrificially etched away the α-Si using $XeF_2$. A total of 1200 minutes of etching was required to completely etch away the sacrificial Si and release the upper electrode.

Imaging, electrical characterization and sensor response High resolution SEM imaging was done at an accelerating voltage of 15.0 kV by the FEI Nova NanoSEM to visually inspect the fabricated device. I-V characteristics of the device were measured using the Keithley 4200A-SCS Semiconductor Parameter Analyzer. Tunneling current measurements were performed in a dark room, where the device under test was placed on a probe-station kept inside an electrically shielded enclosure to ensure low-noise and high fidelity electrical signals. The enclosure had feedthroughs for providing electrical connections to the probe-station. An inlet from a commercially available humidifier with restricted flow was used to control and maintain the humidity levels in the enclosure. An Arduino powered BME280 chip was placed inside the chamber for serial monitoring of the relative humidity levels.

*Conclusion*

We presented the design, fabrication, electrical characterization and working of a temperature compensated tunneling humidity sensor. Sensor response shows a completely reversible reduction of junction resistance of ~five orders of magnitude with a standby DC power consumption of ~0.4 pW when the device is exposed to an increasing RH% of ~ 20-90 RH%. Passive temperature compensation was also achieved using a non-differential swelling design for the polymer patches. Temperature response showed a reduction of ~2.5 times when the device was exposed to a temperature sweep of 25°C – 60°C, which is 0.0025% of the maximum sensor output when exposed to rising levels of humidity. A mathematical model and equivalent electrical circuit was also presented to accurately describe the current conduction mechanism responsible for sensor action. Finally, sensor response dynamics was also investigated and established sorption analytical models were used to describe the time dependent sensor action.

*Author Contributions*

Design, fabrication, device testing, manuscript preparation and proof-reading was done by A.B. R.L assisted in device testing, data acquisition and manuscript proof-reading. H.K was responsible for funding acquisition and providing testing setup. Device's working principle, manuscript preparation, proof-reading and funding acquisition was done by C.M.

*Addition Information*

**Competing Interests:** The authors declare no competing interests.